\documentclass[useAMS,usenatbib]{mn2e}
\usepackage{txfonts}
\usepackage{natbib}
\usepackage[all]{xy}
\usepackage[dvips]{graphicx}

\def\del#1{{}}

\newcommand{\ltsima}{$\; \buildrel < \over \sim \;$}
\newcommand{\lsim}{\lower.5ex\hbox{\ltsima}}
\newcommand{\gtsima}{$\; \buildrel > \over \sim \;$}
\newcommand{\gsim}{\lower.5ex\hbox{\gtsima}}
\newcommand{\bra}{\langle}
\newcommand{\ket}{\rangle}

\newcommand{\dd}{\mathrm{d}}

\title[Lensing constraints on CDM decay]
{Constraints on the decay of dark matter to dark energy from weak lensing bispectrum tomography}
\author[B.M. Sch{\"a}fer, G.A. Caldera-Cabral and R. Maartens]
{Bj{\"o}rn Malte Sch\"afer$^{1,2}$\thanks{e-mail:
bjoern.malte.schaefer@ias.u-psud.fr},
Gabriela Caldera-Cabral$^1$ and Roy Maartens$^1$\\
$^1$ Institute of Cosmology and Gravitation, University of
Portsmouth, Portsmouth PO1 2EG, UK\\
$^2$ Institut d'Astrophysique Spatiale, Universit{\'e} de Paris
XI, b{\^a}timent 120-121, Centre universitaire d'Orsay, 91400
Orsay CEDEX, France}

\begin{document}
\pagerange{\pageref{firstpage}--\pageref{lastpage}} \pubyear{2008}
\maketitle \label{firstpage}

\begin{abstract}
We consider a phenomenological model for a coupling between the dark matter and dark energy fluids and investigate the sensitivity of a weak lensing measurement for constraining the size of this coupling term. Physically, the functional form of the coupling term in our model describes the decay of dark matter into dark energy. We present forecasts for tomographic measurements of the weak shear bispectrum for the DUNE experiment in a Fisher-matrix formalism, where we describe the nonlinearities in structure formation by hyper-extended perturbation theory. Physically, CDM decay tends to increase the growth rate of density perturbations due to higher values for the CDM density at early times, and amplifies the lensing signal because of stronger fluctuations in the gravitational potential. We focus on degeneracies between the dark energy equation of state properties and the CDM decay constant relevant for structure formation and weak lensing. A typical lower bound on the CDM decay time $\simeq7.7/H_0=75.3~\mathrm{Gyr}/h$ which could be provided by DUNE would imply that it would not possible to produce the dark energy content of the universe by CDM decay within the age of the Universe for a constant equation of state parameter of $w$ close to $-1$.
\end{abstract}

\begin{keywords}
cosmology: gravitational lensing, large-scale structure, methods:
analytical
\end{keywords}

\section{Introduction}
Dark energy as a cosmological fluid is evoked for explaining the late-time cosmic acceleration, which has been observed in a number of channels, e.g. the cosmic microwave background (CMB) anisotropies \citep{2003ApJS..148..175S} 
and the integrated Sachs-Wolfe-effect \citep{2003AIPC..666...67B, 2003ApJ...598...97N, 2006PhRvD..74f3520G, 2008arXiv0801.4380G, 2007MNRAS.377.1085R}. The numerical similarity of the values of the dark energy density $\Omega_\phi\simeq0.75$ and the matter density $\Omega_m\simeq0.25$ constitues the {coincidence problem}:
Why is the ratio $\Omega_\phi/\Omega_m=\mathcal{O}(1)$ today, or, why is the Hubble expansion dominated by dark energy after the formation of galaxies and the large-scale structure?

We focus on coupled models of dark energy proposed by \citep[see][]{valiviita,2008arXiv0801.1565B}, where the dark energy (DE) is generated by decay of dark matter (CDM). The CDM decay rate $\Gamma$ is a constant free parameter, in addition to the matter density $\Omega_m$ and the equation of state $w$ of the dark energy. The time evolution
of the matter density $\rho_m$ and the dark energy density $\rho_\phi$ is described by the energy balance equations
\citep{2008arXiv0801.1565B}, which can be written in matrix form as
\begin{equation}
\partial_t\left(\begin{array}{c}\rho_m\\ \rho_\phi\end{array}\right)
+\left(\begin{array}{cc}3H+\Gamma & 0 \\ -\Gamma &
3H(1+w)\end{array}\right) \left(\begin{array}{c}\rho_m\\
\rho_\phi\end{array}\right) = 0, \label{eb}
\end{equation}
with the Friedmann constraint $H^2=\frac{1}{3}\left(\rho_m+\rho_\phi\right)$. Note that the baryons are not coupled to dark energy, since this would be subject to sever constraints from `fifth-force' experiments. We will neglect the baryons (and the radiation) in our analysis, which should not be a serious limitation, however, due to the fact that the baryon fraction $f_b=\Omega_b/\Omega_m\simeq0.16$ has a small numerical value.

We also assume a homogeneous dark energy fluid, even though the decay of CDM in structures will naturally lead to local overdensities of dark energy. If the dark energy sound speed has a value close to 1, the dark energy will diffuse rapidly away from the dark matter structures and constitute a homogeneous fluid. Furthermore, the density fluctuations measured by weak lensing are only mildly overdense such that this assumption may be justified. Furthermore, we will work with the Newtonian Poisson equation and not the full general relativistic expression \citep{2006PhRvD..74d3521O}, which is justified as the dark energy is assumed to be homogeneous.

\citet{2008arXiv0801.1565B} provide a thorough discussion of the background dynamics of such a cosmology in the case where the dark energy is an exponential quintessence field. In order to analyse the growth of structure and weak lensing, we will use a simplified, phenomenological model of dark energy by introducing a parametrised time-variable equation of state. We aim to model a tomographic weak lensing measurement, and thus provide forecasts for parameter constraints and to quantify parameter degeneracies as one can expect from a deep weak lensing survey.

Gravitational lensing, in particular, has been shown to be a very powerful observational probe for investigating the influence of dark energy on  structure formation and the geometry of the universe \citep{1992grle.book.....S, 1999ARA&A..37..127M, 2001PhR...340..291B, 2003ARA&A..41..645R}, even in the nonlinear regime of structure formation \citep{1997ApJ...484..560J, 1997A&A...322....1B, 2001PhRvD..64h3501B}. Lensing data is best used in tomographic measurements for constraining dark energy equation of state properties \citep{1999ApJ...522L..21H, 2002PhRvD..66h3515H, 2003MNRAS.343.1327H, 2003PhRvL..91n1302J}, where one either measures the power spectrum or the bispectrum of a weak lensing quantity \citep{2005A&A...442...69K, 1997MNRAS.286..696S, 2004ApJ...600...17B, 2005PhRvD..72h3001D}. Supplementing the recent paper by \citet{2008arXiv0803.1640L}, who derived lensing bounds on interacting models from weak lensing power spectra, we focus on bispectrum tomography, and we use a more general, albeit phenomenological cosmological model. Bispectra have the advantage that the perturbative treatment is easier to carry out and that they are sensitive on the transition from linear to nonlinear dynamics in structure formation.

After introducing the cosmological model and the peculiarities of gravitational lensing in the decaying CDM models in
Sect.~\ref{sect_homogeneous}, we compute the weak lensing bispectrum and tomographic measurements in Sect.~\ref{sect_tomography}. Fisher-constraints on cosmological parameters are derived in Sect.~\ref{sect_constraints} and the
main results are summarised in Sect.~\ref{sect_summary}. The parameter accuracies are forecast for the weak lensing survey proposed for the Dark UNiverse Explorer\footnote{\tt http://www.dune-mission.net/} (DUNE). For the fiducial model, we take a spatially flat $\Lambda$CDM cosmology with $w=-1$, adiabatic initial conditions and stable CDM ($\Gamma=0$). Specific parameter choices are $H_0=100h \:\mathrm{km}/s/\mathrm{Mpc}$ with $h=0.72$, $\Omega_m=0.25$, $\sigma_8=0.8$ and $n_s=1$.

\section{Growth function and weak lensing}\label{sect_homogeneous}

\subsection{Decaying dark matter}
The decay constant $\Gamma$ will be expressed in units of the Hubble constant $H_0$. We rewrite eqn.~(\ref{eb}) with the scale factor $a$ as independent variable, and introduce the common parameterisation \citep{1997PhRvD..56.4439T, 2001IJMPD..10..213C, 2003MNRAS.346..573L},
\begin{equation}
w(a) = w_0 + (1-a)w_a.
\end{equation}
The Hubble function $H(a)$ is given by solving the differential
equation with solutions for $\rho_m(a)$ and $\rho_\phi(a)$,
\begin{equation}
\frac{\dd H^2}{\dd a} =
-\left(\rho_m(a)+\rho_\phi(a)\left[1+w(a)\right]\right).
\end{equation}
The initial conditions are set at the present epoch $a_0=1$, and
we integrate backwards to $a=10^{-2}$. With the solution of
$H(a)$, the comoving distance is
\begin{equation}
\chi(a) = \int_a^1\:\frac{\dd a}{a^2 H(a)},
\end{equation}
and the density parameters can be defined by normalising $\Omega_m(a)=\rho_m(a)/\rho_\mathrm{crit}(a)$ and
$\Omega_\phi(a)=\rho_\phi(a)/\rho_\mathrm{crit}(a)$ with the critical density $\rho_\mathrm{crit}=3H^2(a)/(8\pi G)$. Table~\ref{table_models} gives an overview of the dark energy models considered in this paper.

\begin{table}\vspace{-0.1cm}
\begin{center}
\begin{tabular}{lccccccl}
\hline\hline model               & $\Omega_m$    & $\sigma_8$    &
$n_s$ & $w_0$         & $w_a$         & $\Gamma$  & CDM
\\
\hline $\Lambda$CDM            &0.25       & 0.8       & 1     &
-1                & 0             & 0     & stable
\\
$\Lambda_\Gamma$CDM &0.25       & 0.8       & 1     & -1
& 0             & $\frac{1}{3}$ & decaying
\\
$\phi$CDM               &0.25       & 0.8       & 1     &
$-\frac{2}{3}$    & $-\frac{1}{3}$    & 0     & stable
\\
$\phi_\Gamma$CDM        &0.25       & 0.8       & 1     &
$-\frac{2}{3}$    & $-\frac{1}{3}$    & $\frac{1}{3}$ & decaying
\\
\hline
\end{tabular}
\end{center}
\caption{Summary of the four primary dark energy models considered
in this paper, where $\Gamma$ is the CDM decay constant.}
\label{table_models}
\end{table}

\begin{figure}
\resizebox{\hsize}{!}{\includegraphics{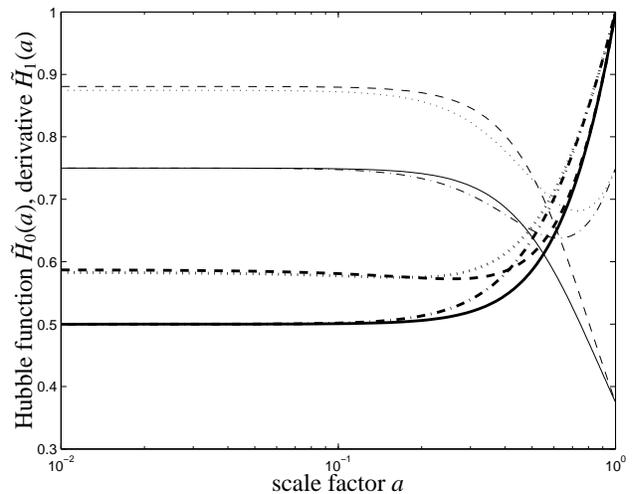}}
\caption{The scaled Hubble function
$\tilde{H}_0(a)=a^{3/2}H(a)/H_0$ (thick lines) and the derivative
$\tilde{H}_1(a)=a^{5/2}\dd H(a)/\dd a/H_0$ (thin lines) in models
with decaying CDM, in comparison to models with stable dark
matter: $\Lambda$CDM (solid line), $\Lambda_\Gamma$CDM with
$\Gamma=\frac{1}{3}$ (dashed line), $\phi$CDM (dash-dotted line),
and $\phi_\Gamma$CDM with $\Gamma=\frac{1}{3}$ (dotted line).}
\label{fig_hubble}
\end{figure}

The Hubble function and its first derivative are depicted in
Fig.~\ref{fig_hubble}, where the SCDM-scaling is divided out,
\begin{equation}
\tilde{H}_n = a^{(2n+3)/2}\frac{\dd^n}{\dd a^n}\frac{H(a)}{H_0}.
\end{equation}
An interesting feature, which will impact on the solution of the
growth equation and on the Poisson equation, is the faster scaling
of the Hubble function with the scale factor $a$ in models with
decay. This arises because the CDM density decreases not only by
redshifting, which would be $\propto a^{-3}$, but also by decay.
In addition, CDM decay causes the Hubble function to vary more
gradually, which similarly can be achieved by introducing a
variable equation of state of the dark energy fluid. Looking at
the scaled first derivative of the Hubble function
$\tilde{H}_1(a)$ illustrates that the interplay between CDM decay
and a variable dark energy equation of state gives rise to
qualitatively new features in the Hubble function at scale factors
$0.2\leq a\leq 0.5$ , compared to dark energy models with stable
CDM.

\subsection{Structure formation}
The linear growth function $D_+(a)$, describing the homogeneous growth of structure, $\delta(\bmath{x},a)=D_+(a)\delta(\bmath{x},1)$ under Newtonian gravity, is obtained by solving the differential equation \citep{1998ApJ...508..483W, 2003MNRAS.346..573L},
\begin{equation} 
\frac{\dd^2}{\dd a^2}D_+(a) + \frac{1}{a}\left(3+\frac{\dd\ln H}{\dd\ln a}\right)\frac{\dd}{\dd a}D_+(a) = 
\frac{3}{2a^2}\Omega_m(a) D_+(a), 
\label{eqn_growth}
\end{equation}
which is still valid in decaying CDM models, as the decay does not affect the overdensity field $\delta(\bmath{x})=(\rho(\bmath{x})-\bra\rho\ket) /\bra\rho\ket$, $\bra\rho\ket=\Omega_m\rho_\mathrm{crit}$. The initial conditions
for decaying CDM models are different compared to those of
standard dark energy models, as the scaling of the Hubble function
as well as the time-evolution of $\Omega_m$ are changed by the
decay. The asymptotic behaviour of the growth equation can be
separated out by assuming $D_+(a)\propto a^\alpha$ with a positive
constant $\alpha$ at early times. Substitution into
eqn.~(\ref{eqn_growth}) yields a quadratic equation for $\alpha$,
which can be solved for an expression of $\alpha$ depending on
$3+\dd\ln H/\dd\ln a$ and $\Omega_m(a)$, evaluated at the initial
time $a_\mathrm{i}$. From that, one obtains the initial conditions
$D_+(a_\mathrm{i})=a_\mathrm{i}^\alpha$ and $\dd D_+/\dd a=\alpha
a_\mathrm{i}^{\alpha-1}$.

\begin{figure}
\resizebox{\hsize}{!}{\includegraphics{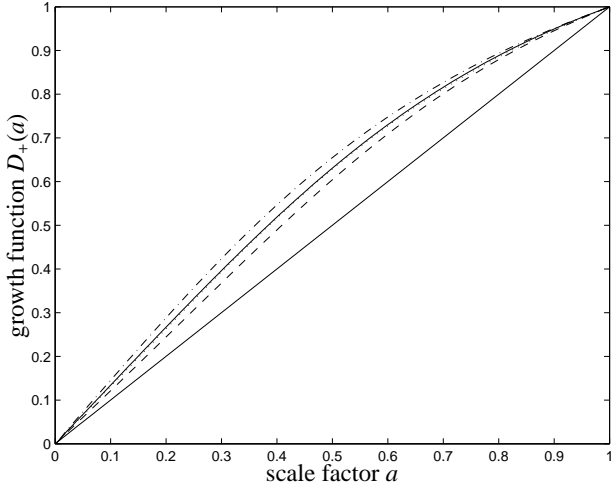}}
\caption{Growth function $D_+(a)$ for four dark energy models:
$\Lambda$CDM (solid line), $\Lambda_\Gamma$CDM with
$\Gamma=\frac{1}{3}$ (dashed line), $\phi$CDM (dash-dotted line),
and $\phi_\Gamma$CDM with $\Gamma=\frac{1}{3}$ (dotted line).
Additionally, the growth function $D_+(a)=a$ for SCDM is plotted
(solid straight line).} \label{fig_growth}
\end{figure}

Fig.~\ref{fig_growth} shows the growth functions $D_+(a)$
(normalised to unity today) in the four cosmologies considered.
Evolving dark energy has the property of suppressing structure
formation at an earlier time, which can be partially compensated
by CDM decay (because the gravitational fields generated by the
overdensity $\delta(\bmath{x})$ are stronger if $\Omega_m(a)$ has
a higher value), hinting at degeneracies between the equation of
state parameters and the CDM decay rate. The $\Lambda$CDM and
$\phi_\Gamma$CDM-models, for example, are almost
indistinguishable. The influence of the two terms $3+\dd\ln
H/\dd\ln a$ and $\Omega_m(a)$ on the evolution of the growth
factor $D_+(a)$ is discussed in
Appendix~\ref{sect_appendix_growth}.

\subsection{Weak lensing}
\begin{figure}
\resizebox{\hsize}{!}{\includegraphics{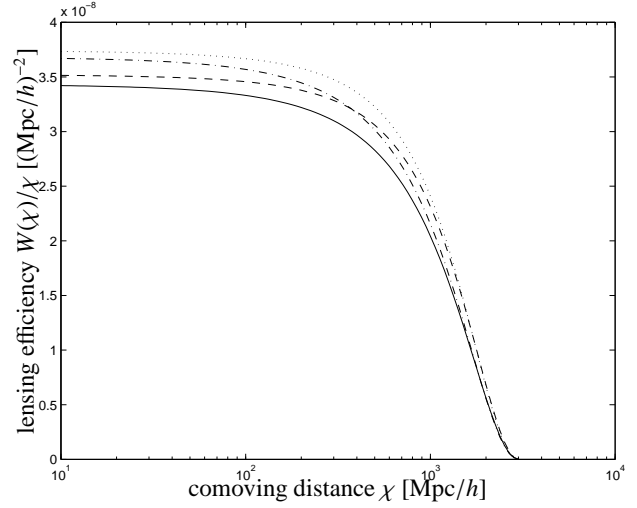}}
\caption{Lensing efficiency functions $W(\chi)/\chi$ for the four
dark energy models: $\Lambda$CDM (solid line), $\Lambda_\Gamma$CDM
with $\Gamma=\frac{1}{3}$ (dashed line), $\phi$CDM (dash-dotted
line), and $\phi_\Gamma$CDM with $\Gamma=\frac{1}{3}$ (dotted
line), without subdivision into tomography bins.}
\label{fig_lensing_efficiency}
\end{figure}

The non-standard scaling of the CDM density with scale factor
makes it impossible to carry out a number of simplifications when
deriving the weak lensing convergence. Substituting the critical
density $\rho_\mathrm{crit}(a)=3H^2(a)/(8\pi G)$ with
$\bra\rho\ket(a) = \Omega_m(a)\rho_\mathrm{crit}(a)$ into the
expression $\Delta\Phi=4\pi G a^2\:\bra\rho\ket(a)\:\delta$ for
the comoving Poisson equation yields
\begin{equation}
\Delta\Phi = \frac{3}{2} a^2 H^2(a)\Omega_m(a)\delta,
\label{eqn_comoving_poisson}
\end{equation}
with explicit functions $H(a)$ and $\Omega_m(a)$. Using this
expression one obtains for the weak lensing convergence
\citep{2001PhR...340..291B}
\begin{equation}
\kappa(\chi) = \frac{3}{2c^2}\int_0^\chi\dd\chi^\prime\: a^2
H^2(a) \Omega_m(a)\:
\frac{\chi\chi^\prime-{\chi^\prime}^2}{\chi}\:\delta.
\end{equation}
Inclusion of the redshift distribution of background galaxies and
reformulating the integration yields:
\begin{equation}
\kappa(\chi) = \frac{3}{2c^2}\int_0^\chi\dd\chi\:a^2 H^2(a)
\Omega_m(a)\:G(\chi)\chi\:\delta,
\end{equation}
where we abbreviated
\begin{equation}
G(\chi) = \int_\chi^{\chi_H}\dd\chi^\prime\: p(z)\frac{\dd
z}{\dd\chi^\prime} \frac{\chi^\prime-\chi}{\chi^\prime},
\end{equation}
with the redshift distribution $p(z)\dd z$ of the lensing
galaxies. From that one finds an analogous expression for the
convergence spectrum \citep{1954ApJ...119..655L},
\begin{equation}
C_\kappa(\ell) = \frac{9}{4c^4}\int_0^{\chi_H}\dd\chi\: G(\chi)^2
\left[a H(a)\right]^4 \Omega_m^2(\chi) D_+^2(a) P(k=\ell/\chi),
\end{equation}
which results from projection of the CDM spectrum $P(k)$. The
standard results are recovered by setting
\begin{equation}
\frac{\Omega_m(a)}{\Omega_m} = \frac{H_0^2}{a^3H^2(a)},
\end{equation}
which is valid only for stable CDM-models. The lensing efficiency
function $W(\chi)$ can be isolated from the above expresssions,
yielding
\begin{equation}
W(\chi) = \frac{3}{2c^2} a^2 H^2(a)\: \Omega_m(a)\: G(\chi)\:
\chi.
\end{equation}
Lensing efficiency functions $W(\chi)/\chi$ for the four exemplary
cosmologies are given in Fig.~\ref{fig_lensing_efficiency}:  At
low redshifts, models with evolving dark energy attain higher
values for $W(\chi)$ compared to models with a cosmological
constant, and at high redshifts, models with decaying CDM have
higher values for $W(\chi)$ compared to models with stable CDM,
with an interesting crossing at a distance of
$\simeq100~\mathrm{Mpc}/h$. This behaviour is caused by the
evolution of $H(a)$ as discussed in the previous section and
indicates that it is possible to increase the redshift at which
the lensing signal originates by increasing $\Gamma$ and by
choosing an equation of state model with mean $w$ close to $-1$.

\begin{figure}
\resizebox{\hsize}{!}{\includegraphics{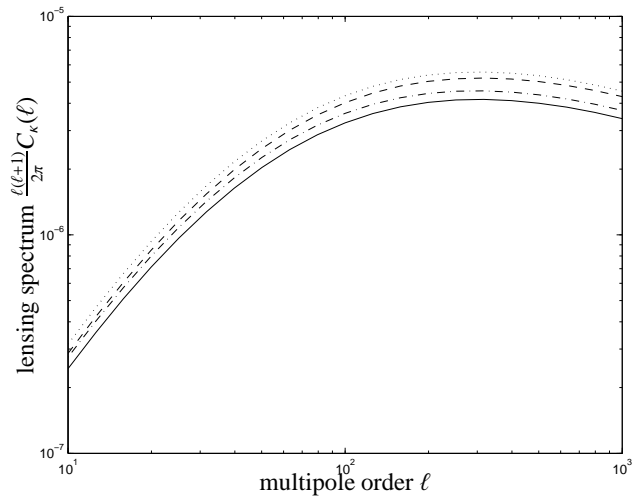}}
\caption{Linear convergence power spectra $C_\kappa(\ell)$ for the
four specified dark energy models: $\Lambda$CDM (solid line),
$\Lambda_\Gamma$CDM with $\Gamma=\frac{1}{3}$ (dashed line),
$\phi$CDM (dash-dotted line), and $\phi_\Gamma$CDM with
$\Gamma=\frac{1}{3}$ (dotted line), without subdivision into
tomography bins.} \label{fig_lensing_spectrum}
\end{figure}

Fig.~\ref{fig_lensing_spectrum} illustrates the impact of CDM
decay on linear convergence power spectra: Models with decaying
CDM exhibit larger values for the power spectrum, which is due to
the fact that in these models the matter density in the past was
higher compared to models with stable CDM, which leads to stronger
gravitational potentials and hence a stronger light deflection,
which is described by the higher values of the lensing efficiency
function $W(\chi)$, with a minor effect from the growth function
$D_+(a)$. This feature might in fact be able to reconcile the high
value for $\sigma_8$ required by weak lensing surveys with the
$\sigma_8$ value following from CMB data, by generating high
values for the lensing signal with a comparatively small value for
$\sigma_8$, if one allows for CDM-decay.

In summary, the lensing signal measures a combination of structure growth $D_+(a)$, coupling strength $\Omega_m(a)$ and geometrical factor $\chi(a)$, which all are influenced by the interaction term in the evolution equations for the dark matter and dark energy density and make up the form of the lensing efficiency $W(\chi)$ and the finally the expression for the spectra $C_\kappa(\ell)$ and $B_\kappa(\ell_1,\ell_2,\ell_3)$. The epoch-varying coupling of the light to the gravitational potential, due to the non-standard Poisson equation eqn.~(\ref{eqn_comoving_poisson}) is an important mechanism, which is not clearly treated in \citet{2008arXiv0803.1640L}. It is worth noting that models with interacting dark fluids constitute a new class of models in this respect, in comparison to models with non-interacting evolving dark energy.

Concerning the numerical evaluation, we generate a lookup table
with 200 values of the relation between scale factor and comoving
distance $a(\chi)$, the Hubble function $H(a)$, the growth
function $D_+(a)$, the matter density parameter $\Omega_m(a)$, and
finally the nonlinear wave vector $k_\mathrm{nlin}$ for a given
cosmological model and use cubic splines for interpolating between
these values. Additionally, values for the tomography lensing
efficiency functions $G_i(\chi)$ for each redshift bin $i$ are
cached, as a function of comoving distance. These measures provide
a significant speed-up in all computations, because we do not have
to reevaluate the densities $\rho_m$ and $\rho_\phi$ for each time
step, to solve for the Hubble function $H(a)$ and the density
parameter $\Omega_m(a)$, and use these results for computing the
growth function $D_+(a)$, the comoving distance $\chi$ and the
lensing efficiency function $G_i(\chi)$.

\section{Weak lensing bispectrum tomography}\label{sect_tomography}
In this section we compile the necessary formulae for a
tomographic measurement of the bispectrum and for a Fisher-matrix
analysis in order to derive constraints on cosmological parameters
including the CDM-decay rate $\Gamma$ and to quantify parameter
degeneracies, especially between $\Gamma$ and the dark energy
properties $w_0$ and $w_a$, from a weak lensing observation.

We choose to consider a measurement of the bispectrum
$B_\kappa(\ell_1,\ell_2,\ell_3)$ rather than a measurement of the
spectrum $C_\kappa(\ell)$ because both yield very similar
constraints on dark energy parameters, as shown by
\citet{2003MNRAS.344..857T, 2004MNRAS.348..897T}, have a high
covariance (which is difficult to quantify as it involves the
computation of a 5-point correlation function) and are by no means
independent measurements, and because of the fact that
perturbation theory for describing the nonlinear evolution of the
cosmic density field is in fact simpler for the 3-point functions
compared to the 2-point functions: A consistent description of the
nonlinear corrections to the 2-point function would involve
contributions from both convergence fields perturbed to first
order, and from one field perturbed to second order while the
other field remains unperturbed. Contrarily, the leading
contribution for the bispectrum involves the perturbation of a
single field to first order, and decomposition of the resulting
4-point function with the Wick-theorem into a product of 2-point
functions.

The most important difference between the spectra and bispectra
concerning the line-of-sight integrations, however, is the fact
that the spectrum is proportional to $W^2(\chi)D_+^2(\chi)$,
whereas the bispectrum  measures different powers of the lensing
efficiency function and the growth rate, namely
$W^3(\chi)D_+^4(\chi)$.

\subsection{Perturbation theory}
For the linear power spectrum $P(k)$, which describes the
fluctuation amplitude of the density field $\delta(\bmath{k})$,
$\bra\delta(\bmath{k})\delta(\bmath{k}^\prime)\ket=
(2\pi)^3\delta_D(\bmath{k}+\bmath{k}^\prime)P(k)$, we make the
ansatz
\begin{equation}
P(k)\propto k^{n_s} T^2(k).
\end{equation}
The shape of the transfer function $T(k)$ is well approximated by
the fitting formula suggested by \citet{1986ApJ...304...15B},
\begin{displaymath}
T(q) = \frac{\ln(1+2.34q)}{2.34q}\left(1+3.89q+(16.1q)^2+(
5.46q)^3+(6.71q)^4\right)^{-\frac{1}{4}},
\end{displaymath}
where the wave vector $k=q\Gamma_s$ is rescaled with the shape
parameter $\Gamma_s$ \citep{1995ApJS..100..281S},
\begin{equation}
\Gamma_s=\Omega_m
h\exp\left(-\Omega_b\left(1+\frac{\sqrt{2h}}{\Omega_m}\right)\right).
\end{equation}
The fluctuation amplitude is normalised to the value $\sigma_8$ on
the scale $R=8~\mathrm{Mpc}/h$,
\begin{equation}
\sigma_R^2 = \frac{1}{2\pi^2}\int\dd k\: k^2 W^2(kR) P(k),
\end{equation}
with a Fourier-transformed spherical top-hat $W(x)=3j_1(x)/x$ as
the filter function. $j_\ell(x)$ denotes the spherical Bessel
function of the first kind of order $\ell$
\citep{1972hmf..book.....A}.

The first order contribution to the bispectrum
$B_\delta(\bmath{k}_1,\bmath{k}_2,\bmath{k}_3)$ of the density
field from nonlinear structure formation is given by
\citep{1984ApJ...277L...5F, 1984ApJ...279..499F,
2003MNRAS.340..580T}:
\begin{equation}
B_\delta(\bmath{k}_1,\bmath{k}_2,\bmath{k}_3) =
\sum_{{(i,j)\in\left\{1,2,3\right\} \atop i\neq j}}
M(\bmath{k}_i,\bmath{k}_j)P_{\mathrm{NL}}(k_i)P_{\mathrm{NL}}(k_j),
\end{equation}
with the classical mode coupling functions,
\begin{equation}
M(\bmath{k}_i,\bmath{k}_j) = \frac{10}{7} +
\left(\frac{k_i}{k_j}+\frac{k_j}{k_i}\right)x + \frac{4}{7}x^2,
\end{equation}
which is replaced by the mode coupling function in hyper-extended
perturbation theory,
\begin{equation}
M(\bmath{k}_i,\bmath{k}_j) = \frac{10}{7} a(k_i)a(k_j) +
b(k_i)b(k_j)\left(\frac{k_i}{k_j}+\frac{k_j}{k_i}\right)x +
\frac{4}{7}c(k_i)c(k_j)x^2,
\end{equation}
where $x =\bmath{k}_i\bmath{k}_j/\left(k_ik_j\right)$ denotes the
cosine of the angle between $\bmath{k}_i$ and $\bmath{k}_j$. The
coefficients $a(k)$, $b(k)$ and $c(k)$ are given by
\citep{1999ApJ...520...35S, 2001MNRAS.325.1312S}:
\begin{eqnarray}
a(k) & = & \frac{1+\sigma_8^{-0.2}(z)\sqrt{0.7Q(n)}\:(q/4)^{n+3.5}}{1+(q/4)^{n+3.5}},\\
b(k) & = & \frac{1+0.4(n+3)q^{n+3}}{1+q^{n+3.5}},\\
c(k) & = &
\frac{1+\frac{4.5}{1.5+(n+3)^4}\:(2q)^{n+3}}{1+(2q)^{n+3.5}},
\end{eqnarray}
where the time evolution of the fluctuation amplitude is given by
the linear growth function, $\sigma_8(z)=D_+(z)\sigma_8$. The wave
vectors are rescaled with the nonlinear wave number, $q\equiv
k/k_{\mathrm{NL}}$. The logarithmic slope of the linear power
spectrum,
\begin{equation}
n(k) = \frac{\dd\ln P(k)}{\dd\ln k}
\end{equation}
can be straightforwardly determined with the polynomial fit to the
transfer function $T(k)$. The nonlinear wave number at scale
factor $a$ is given by the scale at which the variance $\sigma$ of
the density fluctuations becomes unity,
\begin{equation}
\sigma^2=\int_0^{k_\mathrm{NL}}\dd^3k\:D_+^2(z)P(k)=1\rightarrow
4\pi k_\mathrm{NL}^3D_+^2(z)P(k_{\mathrm{NL}})=1.
\end{equation}
Finally, the saturation parameter $Q(n)$ can be computed from the
logarithmic slope $n$ of the linear CDM spectrum,
\begin{equation}
Q(n) = \frac{4-2^n}{1+2^{n+1}}.
\end{equation}
Furthermore, we use the parameterisation proposed by
\citet{2003MNRAS.341.1311S} for the nonlinear CDM spectrum
$P_\mathrm{NL}(k)$ and its slow time evolution, which is
particularly useful as the time evolution is parameterised with
$\Omega_m(a)$.

\subsection{Angular bispectra}
\begin{figure}
\resizebox{\hsize}{!}{\includegraphics{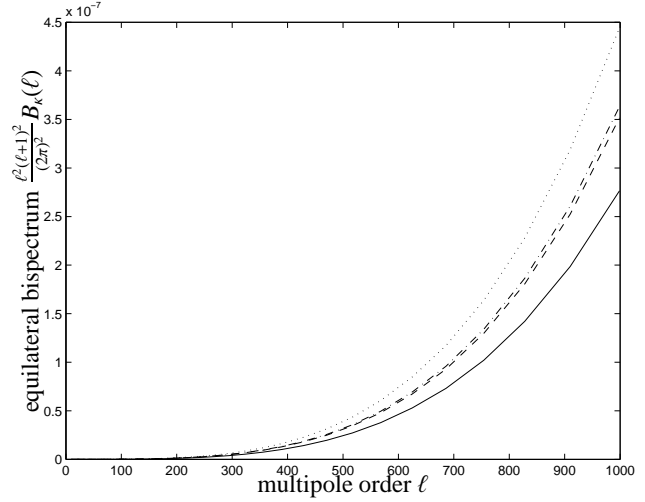}}
\caption{Equilateral convergence bispectra $B_\kappa(\ell)$:
$\Lambda$CDM (solid line), $\Lambda_\Gamma$CDM with
$\Gamma=\frac{1}{3}$ (dashed line), $\phi$CDM (dash-dotted line),
and $\phi_\Gamma$CDM with $\Gamma=\frac{1}{3}$ (dotted line), for
the entire galaxy sample.} \label{fig_equilateral}
\end{figure}

The spherical bispectrum $B_\kappa(\ell_1,\ell_2,\ell_3)$ is
related to the flat-sky bispectrum
$B_\kappa(\bmath{\ell}_1,\bmath{\ell}_2,\bmath{\ell}_3)$ by
\citep{1991ApJ...380....1M, 1992ApJ...388..272K}
\begin{equation}
B_\kappa(\ell_1,\ell_2,\ell_3) \simeq \left(
\begin{array}{ccc}
\ell_1 & \ell_2 & \ell_3\\
0 & 0 & 0
\end{array}
\right) \sqrt{\frac{\prod_{p=1}^3(2\ell_p+1)}{4\pi}}
B_\kappa(\bmath{\ell}_1,\bmath{\ell}_2,\bmath{\ell}_3),
\end{equation}
where
\begin{equation}
\left(\begin{array}{ccc}\ell_1 & \ell_2 & \ell_3\\ 0 & 0 &
0\end{array}\right)^2 = \frac{1}{2}\int_{-1}^{+1}\dd x\:
P_{\ell_1}(x)P_{\ell_2}(x)P_{\ell_3}(x),
\end{equation}
$x=\cos\theta$, denotes the Wigner-$3j$ symbol, which results from
integrating over three Legendre polynomials $P_\ell(x)$. The
Wigner-$3j$ symbol cancels non-admissible configurations which
would violate the triangle inequality
$\left|\ell_i-\ell_j\right|\leq \ell_k \leq \ell_i+\ell_j$
\citep{1972hmf..book.....A}. We use the Limber-equation
\citep{1954ApJ...119..655L} in the flat-sky approximation,
\begin{equation}
B_\kappa(\bmath{\ell}_1,\bmath{\ell}_2,\bmath{\ell}_3) =
\int_0^{\chi_H}\dd\chi\:\frac{1}{\chi^4}\:W^3(\chi)D_+^4(\chi)\:
B_\kappa(\bmath{k}_1,\bmath{k}_2,\bmath{k}_3),
\label{eqn_lensing_bispectrum}
\end{equation}
with $\bmath{k}_p = \bmath{\ell}_p/\chi$, $p=1,2,3$, for
projection of the angular convergence bispectrum $B_\kappa$. The
factorials in the Wigner-$3j$ symbol are evaluated using the
Stirling-approximation for the $\Gamma$-function, $\Gamma(n+1)=n!$
with
\begin{equation}
\Gamma(x)\simeq\sqrt{2\pi}\:\exp(-x)\:x^{x-\frac{1}{2}}
\end{equation}
for $x\gg 1$ \citep{1972hmf..book.....A}. The resulting
approximation for the Wigner-$3j$ symbol overestimates the true
value on average by $\simeq0.4$\% in the relevant $\ell$-range.
Weak lensing convergence bispectra resulting from this procedure
are shown in Fig.~\ref{fig_equilateral} for the equilateral
configuration. As in the case of the convergence spectra
$C_\kappa(\ell)$, the bispectra $B_\kappa(\ell_p)$ attain higher
values in models with decaying CDM compared to those with stable
CDM. This increase amounts to $\simeq10$\%, and is observed
independent of the equation of state of dark energy, where one
observes a change of a few percent in the models with a
cosmological constant relative to those with a varying equation of
state.

\begin{figure}
\resizebox{\hsize}{!}{\includegraphics{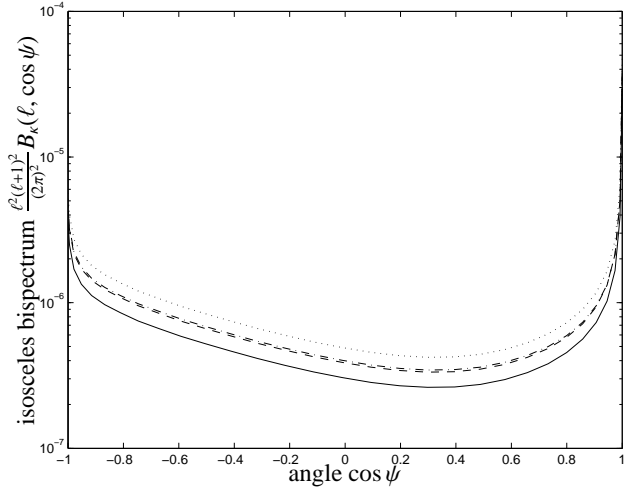}}
\caption{Isosceles convergence bispectra
$B_\kappa(\ell,\cos\psi)$: $\Lambda$CDM (solid line),
$\Lambda_\Gamma$CDM with $\Gamma=\frac{1}{3}$ (dashed line),
$\phi$CDM (dash-dotted line), and $\phi_\Gamma$CDM with
$\Gamma=\frac{1}{3}$ (dotted line), for the entire galaxy sample
as a function of the cosine of the opening angle $\psi$. The
angular scale is fixed to $\ell_1=\ell_2=10^3$.}
\label{fig_isosceles}
\end{figure}

Lensing convergence bispectra for isosceles triangles are given in
Fig.~\ref{fig_isosceles}, for $\ell=10^3$. The plot suggests that
the increase in power observed in decaying CDM models is present
for all configurations, if the background galaxy distribution has
a high average redshift, which is in fact expected for a change of
cosmology on the homogeneous level. Again, typical differences
between models with stable CDM and those with $\Gamma=1/3$ amount
to $\simeq10\%$, irrespective of the opening angle of the
triangle.

\begin{figure}
\resizebox{\hsize}{!}{\includegraphics{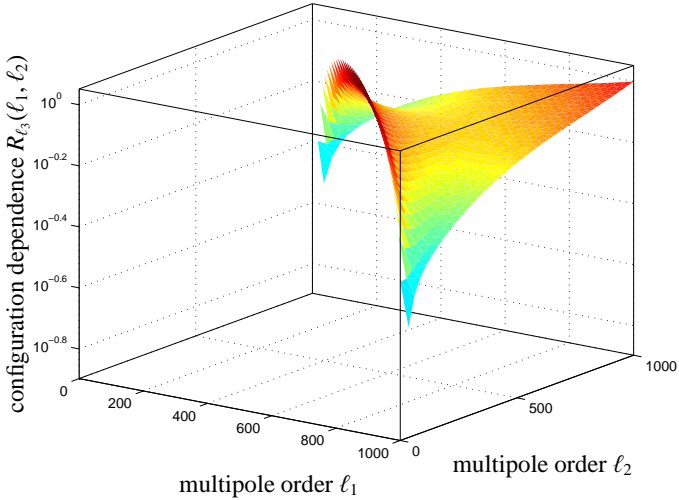}}
\caption{Convergence bispectrum configuration dependence
$R_{\ell_3}(\ell_1,\ell_2)$ for $\ell_3=10^3$, without subdivision
into tomography bins. Non-admissible triangle configurations
violating the inequality
$\left|\ell_i-\ell_j\right|\leq\ell_k\leq\ell_i+\ell_j$ would
occupy the left empty side of the plot.} \label{fig_configuration}
\end{figure}

Fig.~\ref{fig_configuration} illustrates the configuration
dependence of the weak lensing bispectra, for $\ell_3=1000$. In
particular, the plot shows the configuration dependence variable
$R_{\ell_3}(\ell_1,\ell_2)$,
\begin{equation}
R_{\ell_3}(\ell_1,\ell_2) = \frac{\ell_1\ell_2}{\ell_3^2}
\sqrt{\left|\frac{B(\ell_1,\ell_2,\ell_3)}{B(\ell_3,\ell_3,\ell_3)}\right|},
\end{equation}
for $\ell_3=10^3$. Differences in cosmology are cancelled in first
order in this expression, and for that reason the plot only shows
the bispectrum configuration dependence for the fiducial
$\Lambda$CDM model.

\subsection{Bispectrum tomography}
In the case of bispectrum tomography, where the background
galaxies are divided in to two or more redshift bins $(i,j,k)$,
eqn.~(\ref{eqn_lensing_bispectrum}) generalises to
\begin{equation}
B_{(ijk)}(\bmath{\ell}_1,\bmath{\ell}_2,\bmath{\ell}_3) =
\int_0^{\chi_H}\dd\chi\:\frac{W_i(\chi)W_j(\chi)W_k(\chi)}{\chi^4}
D_+^4(\chi) B_\delta(\bmath{k}_1,\bmath{k}_2,\bmath{k}_3),
\end{equation}
with indidual weighting functions
\begin{equation}
W_i(\chi) = \left\{
\begin{array}
{l@{,\:}l}
\frac{3}{2c^2} a^2 H^2(a)\: \Omega_m(a)\: G_i(\chi)\: \chi & \chi\leq\chi_{i+1}\\
0 & \chi>\chi_{i+1},
\end{array}
\right.
\end{equation}
where the lensing-kernel weighted redshift distribution is given
by
\begin{equation}
G_i(\chi)=
\int_{\mathrm{max}(\chi,\chi_i)}^{\chi_{i+1}}\dd\chi^\prime
p(z)\frac{\dd
z}{\dd\chi^\prime}\frac{\chi^\prime-\chi}{\chi^\prime}.
\end{equation}
The tomography bins are chosen to contain equal numbers of
background galaxies, which are distributed in redshift $z$
according to \citep{1995MNRAS.277....1S},
\begin{equation}
p(z)\dd z =
p_0\left(\frac{z}{z_0}\right)^2\exp\left(-\left(\frac{z}{z_0}\right)^\beta\right)\dd
z \quad\mathrm{with}\quad
\frac{1}{p_0}=\frac{z_0}{\beta}\Gamma\left(\frac{3}{\beta}\right),
\end{equation}
where the distribution is characterised by the parameters
$\beta=\frac{3}{2}$ and $z_0=0.64$ (corresponding to a median
redshift of $z_\mathrm{med}=0.9$, c.f.
Appendix~\ref{sect_appendix_zdistro}). We will consider 2-bin
tomography, because any finer subdivision does not necessarily
improve the bounds on cosmological parameters due to the high
covariance of the signal originating from different tomography
bins, as discussed in \citet{2003MNRAS.344..857T,
2004MNRAS.348..897T}.

\section{Parameter constraints}\label{sect_constraints}
We use a Fisher-matrix approach \citep{1997ApJ...480...22T} for
estimating the accuracy of the determination the cosmological
parameters $\Omega_m$, $\sigma_8$, the dark energy properties
$w_0$, $w_a$ and the CDM decay rate $\Gamma$ for the weak lensing
survey to be carried out by the DUNE experiment. The most
important characteristics of DUNE are summarised in
Table~\ref{table_dune}.

\begin{table}\vspace{-0.1cm}
\begin{center}
\begin{tabular}{ccccc}
\hline\hline
$n$                 & $\Delta\Omega$    & $f_\mathrm{sky}$  & $z_0$     & $\sigma_\epsilon$\\
\hline
$4.7\times10^8$     & $2\pi$            & $\frac{1}{2}$     & 0.64      & 0.3\\
\hline
\end{tabular}
\end{center}
\caption{Characteristics of the weak lensing survey of the Dark
UNiverse Explorer (DUNE), in terms of galaxy number $n$ per sterad
(corresponding to a galaxy density of $40/\mathrm{arcmin}^2$),
solid angle $\Delta\Omega$, sky fraction $f_\mathrm{sky}$, survey
depth $z_0$ and intrinsic ellipticity dispersion $\sigma_\epsilon$
of the galaxy sample.} \label{table_dune}
\end{table}

\subsection{Bispectrum covariances}
The observed bispectra $\tilde{B}_{(ijk)}(\ell_1,\ell_2,\ell_3)$
are unbiased estimates of the true bispectra
$B_{(ijk)}(\ell_1,\ell_2,\ell_3)$,
\begin{equation}
\tilde{B}_{(ijk)}(\ell_1,\ell_2,\ell_3) \simeq
B_{(ijk)}(\ell_1,\ell_2,\ell_3),
\end{equation}
because the intrinsic ellipticity distribution
$p(\epsilon)\dd\epsilon$ of the background galaxies is assumed to
be skewless. But $p(\epsilon)\dd\epsilon$ has a finite width
$\sigma_\epsilon$, which impacts on the observed power spectra
$\tilde{C}_{ij}(\ell)$ \citep{1998ApJ...498...26K,
1999ApJ...522L..21H},
\begin{equation}
\tilde{C}_{(ij)}(\ell) = C_{(ij)}(\ell) +
\delta_{ij}\frac{\sigma^2_\epsilon}{n_i}.
\end{equation}
$n_i$ is the number of galaxies per steradian in the tomography
bin $i$,
\begin{equation}
n_i= n\int_{\chi_i}^{\chi_{i+1}}\dd\chi^\prime p(z)\frac{\dd
z}{\dd\chi^\prime},
\end{equation}
with the total number of galaxies per steradian $n$. The
bispectrum covariance for tomography, which is diagonal in
$\ell_p$, $p=1,2,3$, is approximated by
\citep{2000PhRvD..62d3007H, 2003MNRAS.344..857T,
2004MNRAS.348..897T}
\begin{equation}
\mathrm{Cov}\left[B_{(ijk)}(\ell_p)B_{(lmn)}(\ell_p)\right] =
\frac{\Delta(\ell_1,\ell_2,\ell_3)}{f_\mathrm{sky}}
\tilde{C}_{(il)}(\ell_1)\tilde{C}_{(jm)}(\ell_2)\tilde{C}_{(kn)}(\ell_3),
\end{equation}
where the function $\Delta(\ell_p)$ counts the multiplicity of
triangle configurations and is defined as
\begin{equation}
\Delta(\ell_1,\ell_2,\ell_3)= \left\{
\begin{array}
{l@{,\:}l}
6 & \ell_1=\ell_2=\ell_3\\
2 & \ell_i=\ell_j\mathrm{~for~} i\neq j\\
1 & \ell_1\neq\ell_2\neq\ell_3\neq\ell_1.
\end{array}
\right.
\end{equation}
$f_\mathrm{sky}$ denotes the fraction of the observed sky. It is
worth noting that the bispectra $B_{(ijk)}(\ell_1,\ell_2,\ell_3)$
are invariant under permutations of the bin indices. Because of
that, there are 4 independent bispectra for 2-bin tomography,
which can be conveniently indexed with the number $q=2^i+2^j+2^k$.

\subsection{Fisher matrices}
The Fisher matrix $F_{\mu\nu}^{\mathrm{GL}}$ for the parameter
space consisting of $x\in\{\Omega_m,\sigma_8,w_0,w_a,\Gamma,h\}$
is constructed with $\Lambda$CDM as the fiducial cosmological
model. Particularly, for measurements of the weak lensing
bispectrum $B_{(ijk)}(\ell_1,\ell_2,\ell_3)$, one obtains:
\begin{equation}
F_{\mu\nu}^{\mathrm{GL}} =
\sum_{\ell_p=\ell_\mathrm{min}}^{\ell_\mathrm{max}}\sum_{(i,j,k)
\atop (l,m,n)} \frac{\partial B_{(ijk)}(\ell_p)}{\partial
x_\mu}\left(\mathrm{Cov}[B_{(ijk)}
B_{(lmn)}]\right)^{-1}\frac{\partial B_{(lmn)}(\ell_p)}{\partial
x_\nu},
\end{equation}
with $\ell_p\in\{\ell_1,\ell_2,\ell_3\}$. The summation is carried
out with the condition $\ell_1\leq\ell_2\leq\ell_3$, such that
every triangle configuration is counted once, between the limits
$\ell_\mathrm{min}=10^2$ and $\ell_\mathrm{max}=2\times10^3$. The
indices $i,j,k$ and $l,m,n$ run over the tomography bins.

Following \citet{2003MNRAS.344..857T, 2004MNRAS.348..897T}, we use
binned summations in the multipoles $\ell_1$ and $\ell_2$, but
carry out an unbinned summation in $\ell_3$ in order to account
for the vanishing Wigner-$3j$ symbol if $\sum_p\ell_p$ is an odd
number and for the sign change of the Wigner-$3j$ symbol depending
on whether $\sum_p\ell_p~\mathrm{mod}~4$ vanishes or not. Using
$\Delta\ell_1=\Delta\ell_2=25$, the computational load is reduced
by three orders of magnitude, as we have to compute $2\times10^7$
triangles instead of $3\times10^{10}$ triangles, for
$l_\mathrm{max}=2\times10^3$. Using binned summation in two
variables is well justified, given the smooth variation of the
bispectrum illustrated in Fig.~\ref{fig_configuration}. The scale
on which the numerical derivatives are computed corresponds to a
5\% variation for $\Omega_m$, $\sigma_8$ and $h$ and to a 10\%
variation for the dark energy parameters $w_0$, $w_a$ and
$\Gamma$.

We would like to emphasise at this point that our Fisher-analysis
implicitly assumes priors on (i) spatial flatness,
$\Omega_m+\Omega_\phi=1$, which affects all geometrical measures,
(ii) the baryon density $\Omega_b$, which adds a correction to the
shape parameter, and finally (iii) the slope $n_s$ of the CDM
spectrum $P(k)$ for small wave numbers $k$. We neglect the
influence of $\Omega_b$, because the baryons cause only a minor
correction to the shape parameter, and keep $n_s=1$ and
$\Omega_m+\Omega_\phi=1$ fixed, as they are both generic
predictions of inflation and well tested by CMB observations. The
parameter space considered here is motivated by the fact that
$\Omega_m$, $w_0$, $w_a$ and $\Gamma$ all increase the lensing
signal, either by their influence on the growth equation or by
appearing in the Poisson equation, and are naturally degenerate
with $\sigma_8$. As additional priors to the lensing measurement,
we assume $1\sigma$-errors from PLANCK CMB observations with the
numerical values $\Delta h=0.13$ and $\Delta\sigma_8=0.01$
\citep{1999ApJ...518..2E}. The priors are assumed to be diagonal,
$F_{\mu\nu}^\mathrm{CMB}=\delta_{\mu\nu}/\sigma_\mu^2$, and added
to the lensing Fisher matrix $F_{\mu\nu}^\mathrm{GL}$,
\begin{equation}
F_{\mu\nu} = F_{\mu\nu}^\mathrm{GL} + F_{\mu\nu}^\mathrm{CMB},
\end{equation}
because the likelihoods of independent measurements can be
multiplied and their $\chi^2$-functions added.

The diagonal elements of the inverse Fisher matrix give the
Cram{\'e}r-Rao bound on individual parameters,
\begin{equation}
\sigma_\mu = \sqrt{(F^{-1})_{\mu\mu}},
\end{equation}
which are compiled in Table~\ref{table_fisher} for 2-bin weak
lensing tomography, with the enhancement of the measurement by
including CMB-priors on $\sigma_8$ and on $h$. Quite generally,
weak lensing bispectrum tomography with DUNE in conjunction with
CMB priors provides percent errors on $\Omega_m$, $\sigma_8$ and
$h$, errors of the order of 20\% on $w_0$ and $\Gamma$, but a weak
constraint on $w_a$, which is due to the fact that in comparing to
the analysis by \citet{2004MNRAS.348..897T}, the DUNE galaxy
sample has a much lower median redshift (0.9 compared to 1.5) and
hence a weaker lever arm on $w_a$.

The constraint on $\Delta\Gamma\simeq0.13$ translates to a lower
bound on the CDM lifetime of more than 7.7 Hubble times,
corresponding to $t_\Gamma\gsim75.3/h$ Gyr. Limits on the CDM
particle lifetime from lensing measurements of that order of
magnitude may serve to exclude certain particle candidates, but we
should emphasise that we investigate only a single decay channel
with an unknown branching ratio $b$: If CDM has two decay modes
with the probabilities $b$ for decaying into dark energy and $1-b$
for decaying into other products, with corresponding decay rates
$\Gamma_\phi$ (constrained by lensing) and $\Gamma_X$, the total
decay rate is given by $\Gamma=b\Gamma_\phi+(1-b)\Gamma_X$.

\begin{table}\vspace{-0.1cm}
\begin{center}
\begin{tabular}{lll}
\hline\hline
                    & tomography\hphantom{+CMB} & tomography+CMB\\
\hline
$\Omega_m=0.25$     & $\Delta\Omega_m=0.029$        & $\Delta\Omega_m = 0.011$\\
$\sigma_8=0.8$      & $\Delta\sigma_8=0.023$        & $\Delta\sigma_8 = 0.008$\\
$w_0=-1$            & $\Delta w_0=0.443$            & $\Delta w_0 = 0.167$\\
$w_a=0$             & $\Delta w_a=1.956$            & $\Delta w_a = 0.542$\\
$\Gamma=0$          & $\Delta\Gamma=0.187$          & $\Delta\Gamma = 0.130$\\
$h = 0.72$          & $\Delta h=0.085$              & $\Delta h = 0.033$\\
\hline
\end{tabular}
\end{center}
\caption{Expected accuracy on the cosmological parameters from
2-bin weak lensing tomography, with $\Lambda$CDM as the fiducial
cosmology. The second column adds a CMB prior on $\sigma_8$ and
$h$ to the Fisher-matrix.} \label{table_fisher}
\end{table}

\begin{figure*}
\vspace{0.3cm}
\resizebox{0.85\hsize}{!}{\includegraphics{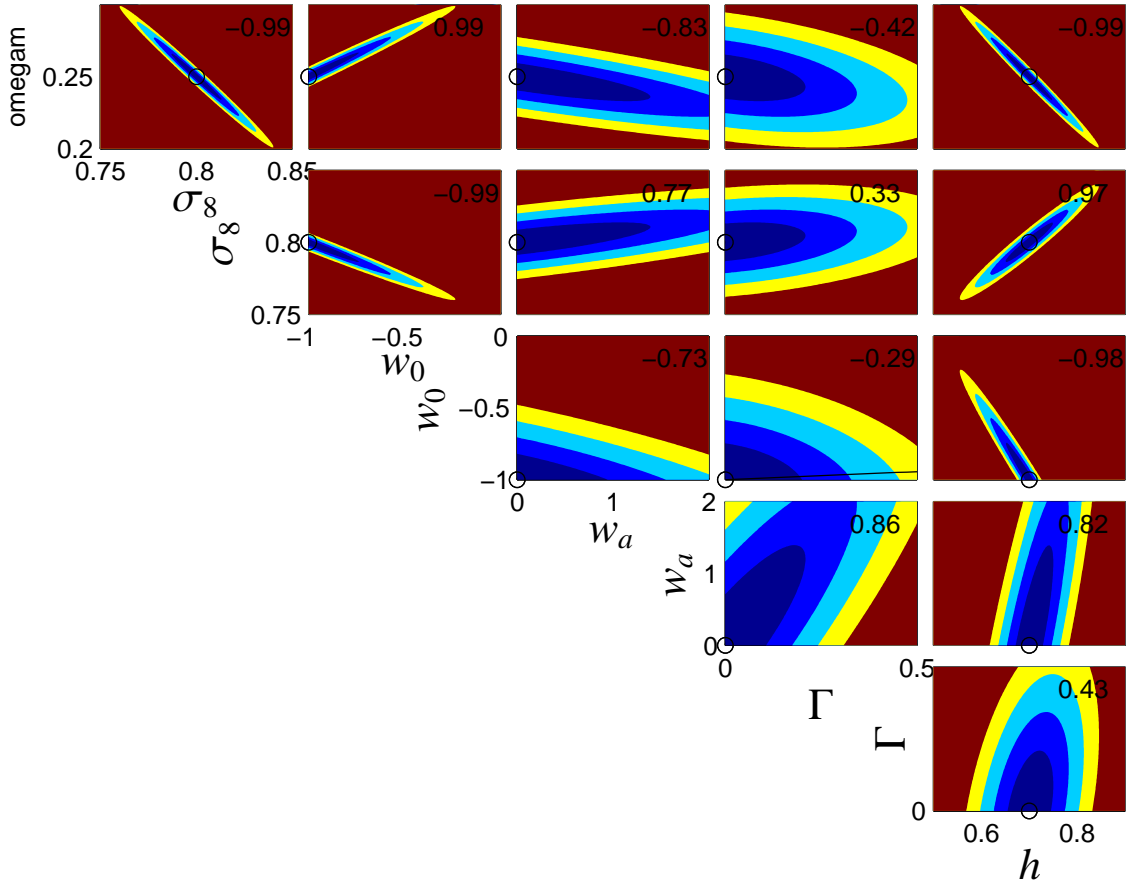}}
\caption{Constraints on cosmological parameters and the CDM decay
rate from the Fisher-matrix analysis from weak lensing tomography.
The choice for the fiducial cosmological model is $\Omega_m=0.25$,
$\sigma_8=0.8$, $w_0=-1$, $w_a=0$, $\Gamma=0$ and $h=0.72$ (marked
by a circle). The ellipses correspond to $1\sigma\ldots4\sigma$.
The correlation coefficient $r_{\mu\nu}$ of the parameter pair
$(x_\mu,x_\nu)$ is given in the upper right corner of each panel
(to two digits), and the the excluded region of parameter space
due to the phantom constraint eqn.~(\ref{eqn_phantom}) is below
the line in the $w_0$-$\Gamma$ plot.} \label{fig_fisher_2d}
\end{figure*}

The $\chi^2$-value for pairs of parameters $(x_\mu,x_\nu)$ can be
computed from the inverse $(F^{-1})_{\mu\nu}$ of the Fisher matrix
\citep{2002ApJ...574....1M},
\begin{equation}
\chi^2 = \left(\begin{array}{c}
\Delta x_\mu \\
\Delta x_\nu
\end{array}\right)^t
\left(\begin{array}{cc}
(F^{-1})_{\mu\mu}   &   (F^{-1})_{\mu\nu}\\
(F^{-1})_{\nu\mu}   &   (F^{-1})_{\nu\nu}
\end{array}\right)^{-1}
\left(\begin{array}{c}
\Delta x_\mu \\
\Delta x_\nu
\end{array}\right),
\label{eqn_chi2}
\end{equation}
where $\Delta x_\mu= x_\mu - x_\mu^{\Lambda\mathrm{CDM}}$.
Specifically, the contour at
\begin{equation}
\Delta\chi^2_n =
-2\ln\:\mathrm{erfc}\left(\frac{n}{\sqrt{2}}\right),
\end{equation}
encloses a fraction of $\mathrm{erf}(n/\sqrt{2})$ of parameter
space, which corresponds to a confidence level of $n\sigma$.
$\mathrm{erf}(x)$ and $\mathrm{erfc}(x)=1-\mathrm{erf}(x)$ are the
error function and the complementary error function, respectively
\citep{1972hmf..book.....A}. The correlation coefficient
$r_{\mu\nu}$ is defined as
\begin{equation}
r_{\mu\nu} =
\frac{(F^{-1})_{\mu\nu}}{\sqrt{(F^{-1})_{\mu\mu}(F^{-1})_{\nu\nu}}},
\end{equation}
and describes the degree of dependence between the parameters
$x_\mu$ and $x_\nu$ by assuming numerical values close to 0 for
independent, and close to 1 for strongly dependent parameters.
Constraints on pairs $(x_\mu,x_\nu)$ of cosmological parameters
from the Fisher-analysis are compiled in Fig.~\ref{fig_fisher_2d},
along with the respective correlation coefficient $r_{\mu\nu}$.

The CDM decay constant $\Gamma$ is negatively correlated with
$\Omega_m$ because a simultaneous decrease of the matter density
today as well as a increase in the decay constant leads to the
same averaged matter density for the lensing signal. The
degeneracies with $w_0$ and $w_a$ are such that $w_0$ has to be
de- and $w_a$ increased in order to have the same lensing signal
in models with non-vanishing $\Gamma$. The positive correlation of
$\Gamma$ with $\sigma_8$ is a projection effect in the
marginalisation driven by the tight constraints between
$\Omega_m$, $w_0$ and $h$, and the classical degeneracies between
$\Omega_m$ and $\sigma_8$ (the power spectrum normalisation and
the strength of the gravitational potentials) and between
$\Omega_m$ and $h$ (the definition of the shape parameter) are
recovered.

We would like to point out that the effective equation of state
parameter $w_\mathrm{eff}$ may cross the boundary towards phantom
models, for which $w<-1$,
\begin{equation}
w_\mathrm{eff} = w - \frac{\Gamma}{3h}\frac{\rho_m}{\rho_\phi}
\stackrel{a=1}{\rightarrow} w_\mathrm{eff} = w -
\frac{\Gamma}{3}\frac{\Omega_m}{\Omega_\phi}, \label{eqn_phantom}
\end{equation}
with $H(a) = H_0 h(a)$, if decay is considered. Therefore, we mark
the forbidden regions of parameter space in the relevant plots.

\begin{figure}
\resizebox{8cm}{!}{\includegraphics{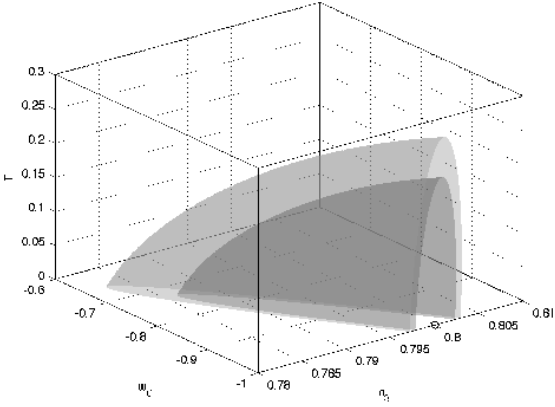}}
\caption{Fisher matrix contraints on the triplet
$(\sigma_8,w_0,\Gamma)$, from weak lensing tomography including a
CMB-prior on $\sigma_8$ and $h$. The ellipsoids correspond to
$1\sigma$ and $2\sigma$ intervals.} \label{fig_fisher_3d1}
\end{figure}

\begin{figure}
\resizebox{8cm}{!}{\includegraphics{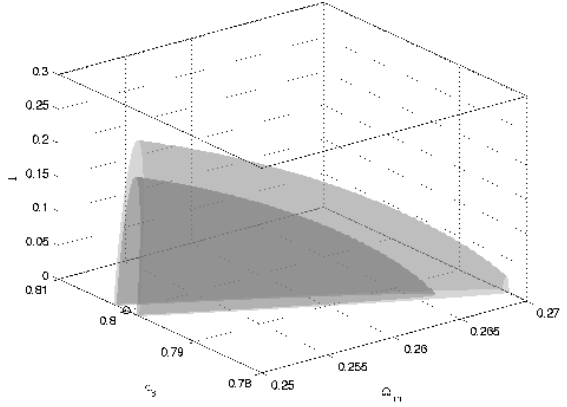}}
\caption{Fisher matrix contraints on the triplet
$(\Omega_m,\sigma_8,\Gamma)$, from weak lensing tomography
including a CMB-prior on $\sigma_8$ and $h$. The ellipsoids
correspond to $1\sigma$ and $2\sigma$ intervals.}
\label{fig_fisher_3d2}
\end{figure}

Generalising eqn.~(\ref{eqn_chi2}) to three parameters
$(x_\mu,x_\nu,x_\rho)$ yields
\begin{equation}
\chi^2 = \left(
\begin{array}{c}
\Delta x_\mu \\
\Delta x_\nu \\
\Delta x_\rho
\end{array}
\right)^t \left(
\begin{array}{ccc}
(F^{-1})_{\mu\mu}           & (F^{-1})_{\mu\nu}         & (F^{-1})_{\mu\rho}        \\
(F^{-1})_{\nu\mu}           & (F^{-1})_{\nu\nu}         & (F^{-1})_{\nu\rho}        \\
(F^{-1})_{\rho\mu}          & (F^{-1})_{\rho\nu}            &
(F^{-1})_{\rho\rho}
\end{array}
\right)^{-1} \left(
\begin{array}{c}
\Delta x_\mu \\
\Delta x_\nu \\
\Delta x_\rho
\end{array}
\right).
\end{equation}
In a 3-dimentional parameter space, the contours at
$\Delta\chi_1^2\simeq 2.6124$ and $\Delta\chi_2^2\simeq4.2222$
correspond to significance levels of $1\sigma$ and $2\sigma$,
respectively. Figs.~\ref{fig_fisher_3d1} and~\ref{fig_fisher_3d2}
summarise simultaneous constraints on the triplets
$(\sigma_8,w_0,\Gamma)$ and $(\Omega_m,\sigma_8,\Gamma)$,
respectively. The increase in lensing signal by either increasing
$\sigma_8$ or the dark energy equation of state $w_0$ is
illustrated in Fig.~\ref{fig_fisher_3d1}, with the weak degeneracy
that models with decaying CDM require higher values of $\sigma_8$,
for lensing at low redshift. Fig.~\ref{fig_fisher_3d2}, on the
contrary, shows that $\sigma_8$ can be decreased in models with
decay if $\Omega_m$ is increased at the same time.

\section{Summary}\label{sect_summary}
In this paper, we investigate the capability of the DUNE
experiment to constrain the decay of dark matter into dark energy
from the observation of the weak convergence bispectrum, in a
tomographic measurement.
\begin{enumerate}
\item{
Our cosmological model with CDM decaying into dark energy may
provide an alternative explanation of the coincidence problem,
i.e. the fact that the cosmic expansion is dominated by dark
energy after structure formation. This is achieved by choosing a
small value of the decay constant $\Gamma$. Decaying CDM naturally
influences the Hubble function by providing dark energy, and by
introducing a faster scaling of the matter density compared to
$\rho\propto a^{-3}$ in models with stable CDM.}

\item{
We have neglected the contribution of baryons, whose density would
just decrease $\propto a^{-3}$, because they are stable particles.
This should not be a serious limitation, however, due to the fact
that the baryon fraction $f_b=\Omega_b/\Omega_m\simeq0.16$ has a
small numerical value. Models with decaying CDM have the
interesting property, however, that the baryon fraction
$f_b=\rho_b/\rho_m$ is monotonically increasing with cosmic time.}

\item{
The growth of structure is influenced in two ways: Firstly by the
non-standard scaling of the density parameter $\Omega_m$ which
causes stronger gravitational potentials in the past in models
with decaying CDM, and secondly by affecting the logarithmic
derivative of the Hubble function, which is smaller in decaying
models. These changes are naturally degenerate with models for
dark energy with a varying equation of state. We work in the limit
of the dark energy sound speed being close to $c$, such that dark
energy can be considered homogeneous even though it originates by
decay from clustered dark matter.}

\item{
Similarly to the growth of structure, the coupling to of light to
the matter in gravitational lensing is affected by the higher
value of $\Omega_m$, which causes the gravitational potential
$\Phi$ induced by the overdensity field $\delta$ to be stronger in
decaying models. With this mechanism, models with decaying CDM can
provide an explanation for the high values of $\sigma_8$ required
by cosmic shear experiments, and can be reconciled with
measurements of $\sigma_8$ from CMB observations.}

\item{Weak lensing bispectrum tomography yields a relative
accuracy on the determination of the dark energy equation of state
parameters $w_0$, $w_a$, the CDM decay rate $\Gamma$ of a tenth
and percent accuracy on the parameters $\Omega_m$, $\sigma_8$, if
CMB priors are included. The CDM decay rate $\Gamma$ is naturally
degenerate with the equation of state parameters $w_0$ and $w_a$.
The Fisher-analysis provides an upper bound on $\Gamma$, or
equivalently, a lower bound on the CDM lifetime,
$t_\Gamma=1/\Gamma> 7.7/H_0\simeq75.3~\mathrm{Gyr}/h$. This limit
might be useful to exclude certain CDM particle candidates,
although we should emphasise here that we investigate a single
decay channel only, and that the branching ratio of this
particular channel would need to be known as well.}

\item{Comparing the lensing constraints on $\Gamma$ to supernova
constraints shows that differences in luminosity distance small:
There is a 0.01\% difference between $\Lambda_\Gamma$CDM and
$\Lambda$CDM and a 1\% difference between $\Phi_\Gamma$CDM and
$\Lambda$CDM at $z=0.1$. Moving to higher redshifts confirms the
trend that the equation of state has a stronger influence on the
luminosity distance than the CDM decay rate. At $z=1$, one
observes a 1\% difference between $\Lambda_\Gamma$CDM and
$\Lambda$CDM, but 7\% difference between the stable dark energy
models $\Phi_\Gamma$CDM and $\Lambda$CDM.}

\item{It is worth noting that models with decaying CDM are genuinely
different from dark energy models concerning structure growth and
observations which use gravitational interaction such as
gravitational lensing. While it is always possible to construct an
equation of state $w(a)$ for a dark energy model with stable CDM
that gives the identical Hubble function as a model with decaying
CDM, the evolution of the density parameter $\Omega_m(a)$ and the
growth function $D_+(a)$ breaks this degeneracy which would be
different in the two cases. For that reason, the combination of
probes of cosmic structure growth and the expansion history are
able to distinguish between the two families of models, which
would not be possible with e.g. supernova observations alone.}
\end{enumerate}

In addition, we plan to provide prospective constraints from the integrated Sachs-Wolfe effect, which is promising as it originates at higher redshifts, and directly measures the derivative $\dd\ln\rho_m(a)/\dd\ln a$, which is $\neq -3$ in decaying CDM models. In summary we would like to stress that CDM-decay would be an elegant solution to the coincidence problem, and that its central parameter $\Gamma$ is well measurable by future lensing surveys.

\section*{Acknowledgements}
We would like to thank David Bacon, Rob Crittenden, and Lukas Hollenstein for valuable comments, and Alexandre Refregier for providing the characteristics of the DUNE survey. BMS acknowledges support from an STFC postdoctoral fellowship. GACC is supported by the Programme Alban, the European Union Programme of High Level
Scholarships for Latin America, scholarship No. E06D103604MX and the Mexican National Council for Science and Technology, CONACYT, scholarship No. 192680. The work of RM is supported by STFC.

\bibliography{bibtex/aamnem,bibtex/references}

\begin{thebibliography}{}

\bibitem[\protect\citeauthoryear{{Abramowitz} \& {Stegun}}{{Abramowitz} \&
  {Stegun}}{1972}]{1972hmf..book.....A}
{Abramowitz} M.,  {Stegun} I.~A.,  1972, {Handbook of Mathematical Functions}.
Handbook of Mathematical Functions, New York: Dover, 1972

\bibitem[\protect\citeauthoryear{{Bardeen}, {Bond}, {Kaiser} \&
  {Szalay}}{{Bardeen} et~al.}{1986}]{1986ApJ...304...15B}
{Bardeen} J.~M.,  {Bond} J.~R.,  {Kaiser} N.,    {Szalay} A.~S.,  1986, \apj,
  304, 15

\bibitem[\protect\citeauthoryear{{Bartelmann} \& {Schneider}}{{Bartelmann} \&
  {Schneider}}{2001}]{2001PhR...340..291B}
{Bartelmann} M.,  {Schneider} P.,  2001, \physrep, 340, 291

\bibitem[\protect\citeauthoryear{{Benabed} \& {Bernardeau}}{{Benabed} \&
  {Bernardeau}}{2001}]{2001PhRvD..64h3501B}
{Benabed} K.,  {Bernardeau} F.,  2001, \prd, 64, 083501

\bibitem[\protect\citeauthoryear{{Bernardeau}, {van Waerbeke} \&
  {Mellier}}{{Bernardeau} et~al.}{1997}]{1997A&A...322....1B}
{Bernardeau} F.,  {van Waerbeke} L.,    {Mellier} Y.,  1997, \aap, 322, 1

\bibitem[\protect\citeauthoryear{{Bernstein} \& {Jain}}{{Bernstein} \&
  {Jain}}{2004}]{2004ApJ...600...17B}
{Bernstein} G.,  {Jain} B.,  2004, \apj, 600, 17

\bibitem[\protect\citeauthoryear{{Boehmer}, {Caldera-Cabral}, {Lazkoz} \&
  {Maartens}}{{Boehmer} et~al.}{2008}]{2008arXiv0801.1565B}
{Boehmer} C.~G.,  {Caldera-Cabral} G.,  {Lazkoz} R.,    {Maartens} R.,  2008,
  ArXiv 0801.1565, 801

\bibitem[\protect\citeauthoryear{{Boughn} \& {Crittenden}}{{Boughn} \&
  {Crittenden}}{2003}]{2003AIPC..666...67B}
{Boughn} S.~P.,  {Crittenden} R.~G.,  2003, in {Holt} S.~H.,  {Reynolds} C.~S.,
   eds, The Emergence of Cosmic Structure Vol.~666 of American Institute of
  Physics Conference Series, {The Absence of the Integrated Sachs-Wolfe Effect:
  Constraints on a Cosmological Constant}.
pp 67--70

\bibitem[\protect\citeauthoryear{{Chevallier} \& {Polarski}}{{Chevallier} \&
  {Polarski}}{2001}]{2001IJMPD..10..213C}
{Chevallier} M.,  {Polarski} D.,  2001, International Journal of Modern Physics
  D, 10, 213

\bibitem[\protect\citeauthoryear{{Dodelson} \& {Zhang}}{{Dodelson} \&
  {Zhang}}{2005}]{2005PhRvD..72h3001D}
{Dodelson} S.,  {Zhang} P.,  2005, \prd, 72, 083001

\bibitem[\protect\citeauthoryear{{Eisenstein}, {Hu} \& {Tegmark}}{{Eisenstein}
  et~al.}{1999}]{1999ApJ...518..2E}
{Eisenstein} D.~J.,  {Hu} W.,    {Tegmark} M.,  1999, \apj, 518, 2

\bibitem[\protect\citeauthoryear{{Fry}}{{Fry}}{1984a}]{1984ApJ...277L...5F}
{Fry} J.~N.,  1984a, \apjl, 277, L5

\bibitem[\protect\citeauthoryear{{Fry}}{{Fry}}{1984b}]{1984ApJ...279..499F}
{Fry} J.~N.,  1984b, \apj, 279, 499

\bibitem[\protect\citeauthoryear{{Giannantonio}, {Crittenden}, {Nichol},
  {Scranton}, {Richards}, {Myers}, {Brunner}, {Gray}, {Connolly} \&
  {Schneider}}{{Giannantonio} et~al.}{2006}]{2006PhRvD..74f3520G}
{Giannantonio} T.,  {Crittenden} R.~G.,  {Nichol} R.~C.,  {Scranton} R.,
  {Richards} G.~T.,  {Myers} A.~D.,  {Brunner} R.~J.,  {Gray} A.~G.,
  {Connolly} A.~J.,    {Schneider} D.~P.,  2006, \prd, 74, 063520

\bibitem[\protect\citeauthoryear{{Giannantonio}, {Scranton}, {Crittenden},
  {Nichol}, {Boughn}, {Myers} \& {Richards}}{{Giannantonio}
  et~al.}{2008}]{2008arXiv0801.4380G}
{Giannantonio} T.,  {Scranton} R.,  {Crittenden} R.~G.,  {Nichol} R.~C.,
  {Boughn} S.~P.,  {Myers} A.~D.,    {Richards} G.~T.,  2008, ArXiv 0801.4380,
  801

\bibitem[\protect\citeauthoryear{{Heavens}}{{Heavens}}{2003}]{2003MNRAS.343.13%
27H}
{Heavens} A.,  2003, \mnras, 343, 1327

\bibitem[\protect\citeauthoryear{{Hu}}{{Hu}}{1999}]{1999ApJ...522L..21H}
{Hu} W.,  1999, \apjl, 522, L21

\bibitem[\protect\citeauthoryear{{Hu}}{{Hu}}{2000}]{2000PhRvD..62d3007H}
{Hu} W.,  2000, \prd, 62, 043007

\bibitem[\protect\citeauthoryear{{Hu}}{{Hu}}{2002}]{2002PhRvD..66h3515H}
{Hu} W.,  2002, \prd, 66, 083515

\bibitem[\protect\citeauthoryear{{Jain} \& {Seljak}}{{Jain} \&
  {Seljak}}{1997}]{1997ApJ...484..560J}
{Jain} B.,  {Seljak} U.,  1997, \apj, 484, 560

\bibitem[\protect\citeauthoryear{{Jain} \& {Taylor}}{{Jain} \&
  {Taylor}}{2003}]{2003PhRvL..91n1302J}
{Jain} B.,  {Taylor} A.,  2003, Physical Review Letters, 91, 141302

\bibitem[\protect\citeauthoryear{{Kaiser}}{{Kaiser}}{1992}]{1992ApJ...388..272%
K}
{Kaiser} N.,  1992, \apj, 388, 272

\bibitem[\protect\citeauthoryear{{Kaiser}}{{Kaiser}}{1998}]{1998ApJ...498...26%
K}
{Kaiser} N.,  1998, \apj, 498, 26

\bibitem[\protect\citeauthoryear{{Kilbinger} \& {Schneider}}{{Kilbinger} \&
  {Schneider}}{2005}]{2005A&A...442...69K}
{Kilbinger} M.,  {Schneider} P.,  2005, \aap, 442, 69

\bibitem[\protect\citeauthoryear{{La Vacca} \& {Colombo}}{{La Vacca} \&
  {Colombo}}{2008}]{2008arXiv0803.1640L}
{La Vacca} G.,  {Colombo} L.~P.~L.,  2008, ArXiv e-prints, 803

\bibitem[\protect\citeauthoryear{{Limber}}{{Limber}}{1954}]{1954ApJ...119..655%
L}
{Limber} D.~N.,  1954, \apj, 119, 655

\bibitem[\protect\citeauthoryear{{Linder} \& {Jenkins}}{{Linder} \&
  {Jenkins}}{2003}]{2003MNRAS.346..573L}
{Linder} E.~V.,  {Jenkins} A.,  2003, \mnras, 346, 573

\bibitem[\protect\citeauthoryear{{Matsubara} \& {Szalay}}{{Matsubara} \&
  {Szalay}}{2002}]{2002ApJ...574....1M}
{Matsubara} T.,  {Szalay} A.~S.,  2002, \apj, 574, 1

\bibitem[\protect\citeauthoryear{{Mellier}}{{Mellier}}{1999}]{1999ARA&A..37..1%
27M}
{Mellier} Y.,  1999, \araa, 37, 127

\bibitem[\protect\citeauthoryear{{Miralda-Escude}}{{Miralda-Escude}}{1991}]{19%
91ApJ...380....1M}
{Miralda-Escude} J.,  1991, \apj, 380, 1

\bibitem[\protect\citeauthoryear{{Nolta}, {Devlin}, {Dorwart}, {Miller},
  {Page}, {Puchalla}, {Torbet} \& {Tran}}{{Nolta}
  et~al.}{2003}]{2003ApJ...598...97N}
{Nolta} M.~R.,  {Devlin} M.~J.,  {Dorwart} W.~B.,  {Miller} A.~D.,  {Page}
  L.~A.,  {Puchalla} J.,  {Torbet} E.,    {Tran} H.~T.,  2003, \apj, 598, 97

\bibitem[\protect\citeauthoryear{{Olivares}, {Atrio-Barandela} \&
  {Pav{\'o}n}}{{Olivares} et~al.}{2006}]{2006PhRvD..74d3521O}
{Olivares} G.,  {Atrio-Barandela} F.,    {Pav{\'o}n} D.,  2006, \prd, 74,
  043521

\bibitem[\protect\citeauthoryear{{Rassat}, {Land}, {Lahav} \&
  {Abdalla}}{{Rassat} et~al.}{2007}]{2007MNRAS.377.1085R}
{Rassat} A.,  {Land} K.,  {Lahav} O.,    {Abdalla} F.~B.,  2007, \mnras, 377,
  1085

\bibitem[\protect\citeauthoryear{{Refregier}}{{Refregier}}{2003}]{2003ARA&A..4%
1..645R}
{Refregier} A.,  2003, \araa, 41, 645

\bibitem[\protect\citeauthoryear{{Schneider} \& {Bartelmann}}{{Schneider} \&
  {Bartelmann}}{1997}]{1997MNRAS.286..696S}
{Schneider} P.,  {Bartelmann} M.,  1997, \mnras, 286, 696

\bibitem[\protect\citeauthoryear{{Schneider}, {Ehlers} \& {Falco}}{{Schneider}
  et~al.}{1992}]{1992grle.book.....S}
{Schneider} P.,  {Ehlers} J.,    {Falco} E.~E.,  1992, {Gravitational Lenses}.
Springer-Verlag Berlin Heidelberg New York.

\bibitem[\protect\citeauthoryear{{Scoccimarro} \& {Couchman}}{{Scoccimarro} \&
  {Couchman}}{2001}]{2001MNRAS.325.1312S}
{Scoccimarro} R.,  {Couchman} H.~M.~P.,  2001, \mnras, 325, 1312

\bibitem[\protect\citeauthoryear{{Scoccimarro} \& {Frieman}}{{Scoccimarro} \&
  {Frieman}}{1999}]{1999ApJ...520...35S}
{Scoccimarro} R.,  {Frieman} J.~A.,  1999, \apj, 520, 35

\bibitem[\protect\citeauthoryear{{Smail}, {Hogg}, {Blandford}, {Cohen}, {Edge}
  \& {Djorgovski}}{{Smail} et~al.}{1995}]{1995MNRAS.277....1S}
{Smail} I.,  {Hogg} D.~W.,  {Blandford} R.,  {Cohen} J.~G.,  {Edge} A.~C.,
  {Djorgovski} S.~G.,  1995, \mnras, 277, 1

\bibitem[\protect\citeauthoryear{{Smith}, {Peacock}, {Jenkins}, {White},
  {Frenk}, {Pearce}, {Thomas}, {Efstathiou} \& {Couchman}}{{Smith}
  et~al.}{2003}]{2003MNRAS.341.1311S}
{Smith} R.~E.,  {Peacock} J.~A.,  {Jenkins} A.,  {White} S.~D.~M.,  {Frenk}
  C.~S.,  {Pearce} F.~R.,  {Thomas} P.~A.,  {Efstathiou} G.,    {Couchman}
  H.~M.~P.,  2003, \mnras, 341, 1311

\bibitem[\protect\citeauthoryear{{Spergel}, {Verde}, {Peiris}, {Komatsu},
  {Nolta}, {Bennett}, {Halpern}, {Hinshaw}, {Jarosik}, {Kogut}, {Limon},
  {Meyer}, {Page}, {Tucker}, {Weiland}, {Wollack} \& {Wright}}{{Spergel}
  et~al.}{2003}]{2003ApJS..148..175S}
{Spergel} D.~N.,  {Verde} L.,  {Peiris} H.~V.,  {Komatsu} E.,  {Nolta} M.~R.,
  {Bennett} C.~L.,  {Halpern} M.,  {Hinshaw} G.,  {Jarosik} N.,  {Kogut} A.,
  {Limon} M.,  {Meyer} S.~S.,  {Page} L.,  {Tucker} G.~S.,  {Weiland} J.~L.,
  {Wollack} E.,    {Wright} E.~L.,  2003, \apjs, 148, 175

\bibitem[\protect\citeauthoryear{{Sugiyama}}{{Sugiyama}}{1995}]{1995ApJS..100.%
.281S}
{Sugiyama} N.,  1995, \apjs, 100, 281

\bibitem[\protect\citeauthoryear{{Takada} \& {Jain}}{{Takada} \&
  {Jain}}{2003a}]{2003MNRAS.340..580T}
{Takada} M.,  {Jain} B.,  2003a, \mnras, 340, 580

\bibitem[\protect\citeauthoryear{{Takada} \& {Jain}}{{Takada} \&
  {Jain}}{2003b}]{2003MNRAS.344..857T}
{Takada} M.,  {Jain} B.,  2003b, \mnras, 344, 857

\bibitem[\protect\citeauthoryear{{Takada} \& {Jain}}{{Takada} \&
  {Jain}}{2004}]{2004MNRAS.348..897T}
{Takada} M.,  {Jain} B.,  2004, \mnras, 348, 897

\bibitem[\protect\citeauthoryear{{Tegmark}, {Taylor} \& {Heavens}}{{Tegmark}
  et~al.}{1997}]{1997ApJ...480...22T}
{Tegmark} M.,  {Taylor} A.~N.,    {Heavens} A.~F.,  1997, \apj, 480, 22

\bibitem[\protect\citeauthoryear{{Turner} \& {White}}{{Turner} \&
  {White}}{1997}]{1997PhRvD..56.4439T}
{Turner} M.~S.,  {White} M.,  1997, \prd, 56, 4439

\bibitem[\protect\citeauthoryear{{Valiviita}}{{Valiviita}}{2008}]{valiviita}
{Valiviita} J.,  2008, {\prd}.
in preparation

\bibitem[\protect\citeauthoryear{{Wang} \& {Steinhardt}}{{Wang} \&
  {Steinhardt}}{1998}]{1998ApJ...508..483W}
{Wang} L.,  {Steinhardt} P.~J.,  1998, \apj, 508, 483

\end{thebibliography}
\bibliographystyle{mn2e}

\appendix

\section{Growth function}\label{sect_appendix_growth}
\begin{figure}
\resizebox{\hsize}{!}{\includegraphics{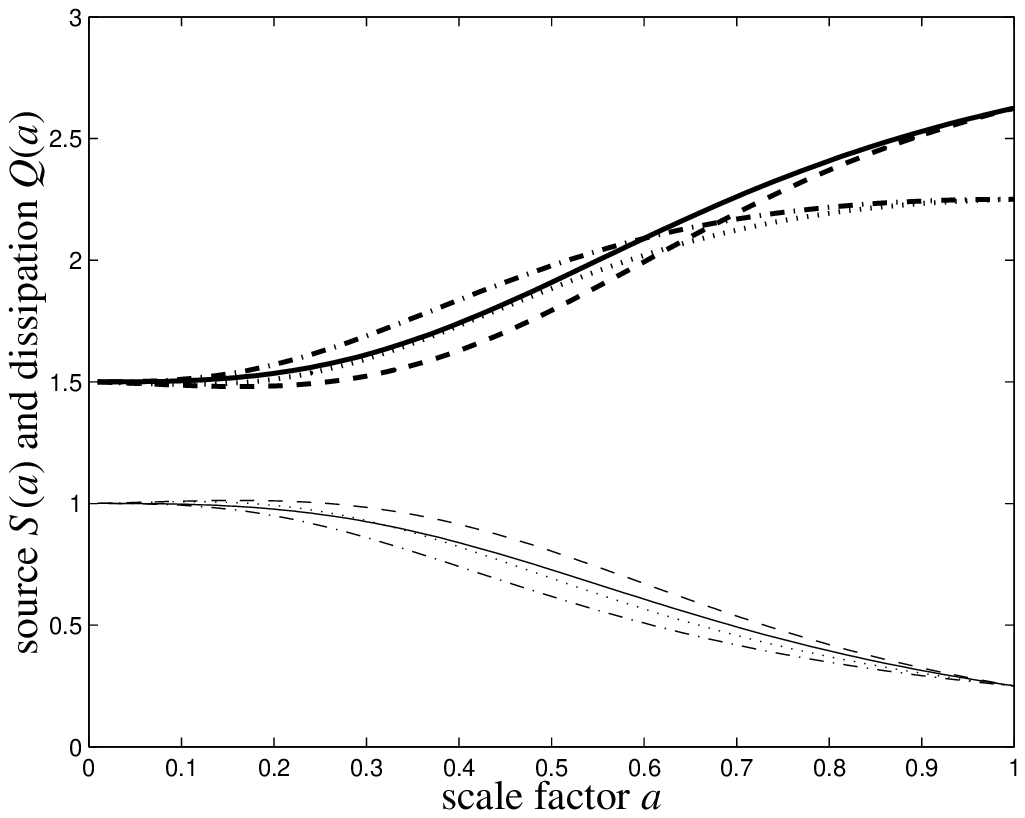}}
\caption{Comparison of the dissipation term $Q(a)=3+\dd\ln
H/\dd\ln a$ (thick lines) and the source term $S(a)=\Omega_m(a)$
(thin lines) in the growth equation, for  $\Lambda$CDM (solid
line), $\Lambda_\Gamma$CDM with $\Gamma=\frac{1}{3}$ (dashed
line), $\phi$CDM (dash-dotted line), and $\phi_\Gamma$CDM with
$\Gamma=\frac{1}{3}$ (dotted line).} \label{fig_growth_terms}
\end{figure}

The growth function $D_+(a)$ is a solution to the differential
equation
\begin{equation}
\frac{\dd^2}{\dd a^2}D_+(a) + \frac{1}{a}\left(3+\frac{\dd\ln
H}{\dd\ln a}\right)\frac{\dd}{\dd a}D_+(a) =
\frac{3}{2a^2}\Omega_m(a)D_+(a).
\end{equation}
Fig.~\ref{fig_growth_terms} compares the source term
$S(a)=\Omega_m(a)$ with the dissipation term $Q(a)=2-q=3+\dd\ln
H/\dd\ln a$ for the exemplary cosmologies. Dark energy with a
variable equation of state increases the dissipation term and
decreases the source term of the growth equation, and suppresses
growth relative to $\Lambda$CDM. CDM decay is able to compensate
this effect, by enhancing growth due to decreasing the dissipation
term $Q(a)$ even below the value of 1.5, and a slower increase at
late times. At the same time, the matter density remains closer to
the critical density for a longer period of time, until the CDM
decay causes it to decrease to the current value.

The near-degeneracy of the growth functions $D_+(a)$ in
$\Lambda$CDM and $\phi_\Gamma$CDM are explained by the fact that
they show very similary evolutions of $\Omega_m(a)$, and up to
$z=1$, the damping term $Q(a)$ is almost identical between the two
models. At smaller redshifts, the damping as well as the source
term are both smaller compared to $\Lambda$CDM, with a
compensating effect on the evolution of $D_+(a)$.

\section{Redshift distribution}\label{sect_appendix_zdistro}
\begin{figure}
\resizebox{\hsize}{!}{\includegraphics{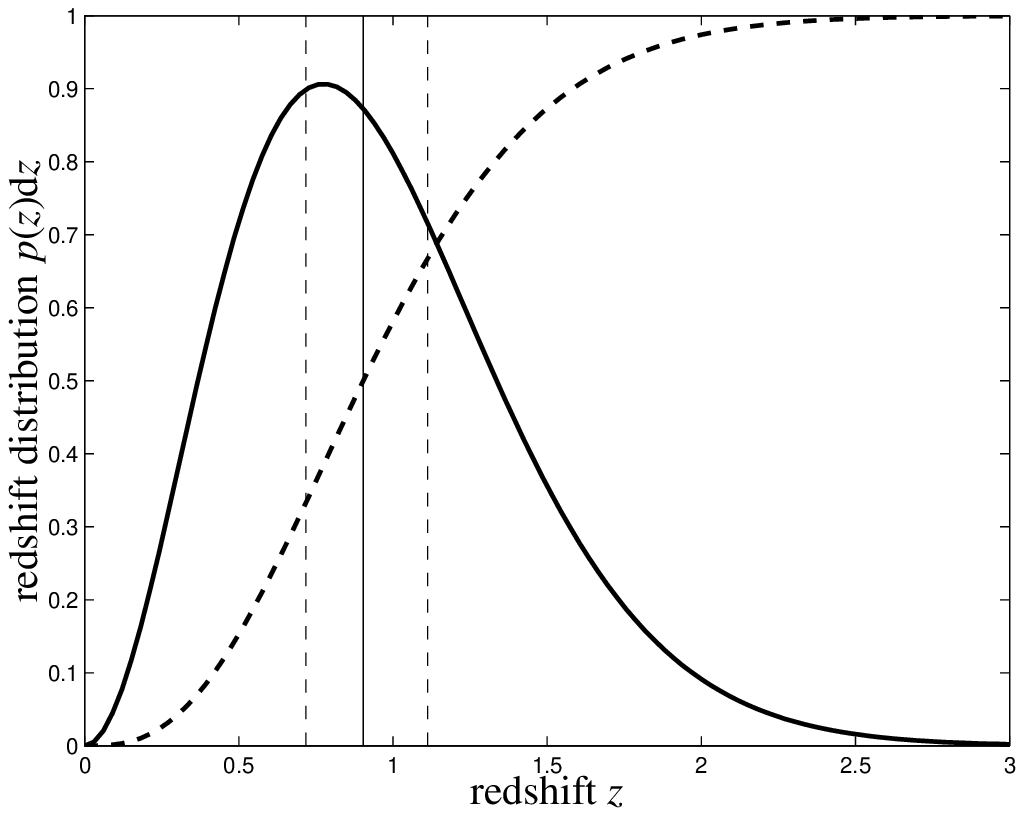}}
\caption{Differential distribution $p(z)\dd z$ (thick solid line)
and cumulative distribution $P(z)=\int_0^z\dd z^\prime
p(z^\prime)$ (thick dashed line) of the lensed galaxies observed
by DUNE, with redshift bins for 2-bin (thin solid line) and 3-bin
(thin dashed line) weak lensing tomography.} \label{fig_zdistro}
\end{figure}

The unit-normalised redshift distribution $p(z)\dd z$
\citep{1995MNRAS.277....1S},
\begin{equation}
p(z)\dd z =
p_0\left(\frac{z}{z_0}\right)^2\exp\left(-\left(\frac{z}{z_0}\right)^\beta\right)\dd
z \quad\mathrm{with}\quad
\frac{1}{p_0}=\frac{z_0}{\beta}\Gamma\left(\frac{3}{\beta}\right),
\end{equation}
the cumulative distribution $P(z)$,
\begin{equation}
P(z)=\int_0^z\dd z^\prime p(z^\prime),
\end{equation}
and the redshift boundaries for 2-bin ($z=0.90$) and 3-bin
($z=0.72$ and $1.11$) tomography are shown in
Fig.~\ref{fig_zdistro}. The parameters are $z_0=0.64$
(corresponding to a median redshift of $z_\mathrm{med}=0.9$) and
$\beta=\frac{3}{2}$. These bins are chosen in such a way that
there are equal number of background galaxies in each tomography
bin. This has the consequence that the noise contribution
$\sigma_\epsilon^2/n_i$ to the spectral measurements from each bin
is equal. Although convenient, this choice of redshift binning is
by no means optimised for providing the maximal accuracy in the
measurement of cosmological parameters. It should be possible,
however, to perform an iterative procedure with a coarse
determination to begin with, and a subsequent choice of tomography
bins which maximises the accuracy.

\bsp

\label{lastpage}

\end{document}